\documentclass[iop]{emulateapj}
\usepackage{natbib}
\usepackage{epsfig}
\usepackage{graphics}
\usepackage{lscape}
\usepackage{verbatim}

\newcommand{\dg}{$^{\circ}$}

\newcommand{\hi}{H{\sc i}}

\newcommand{\hii}{H{\sc ii}}
\newcommand{\amin}{$`$}
\newcommand{\asec}{$``$}
\newcommand{\kms}{\mbox{ km s$^{-1}$}}
\newcommand{\kkms}{\mbox{K~km~s$^{-1}$}}
\newcommand{\lesim}{\raisebox{-0.4ex}{$\stackrel{<}{\scriptstyle\sim}$}}
\newcommand{\msol}{M$_{\odot}$}
\newcommand{\cp}{{\sc cloudprops}}
\newcommand{\mvir}{M$_{VIR}$}
\newcommand{\ico}{I$_{CO}$}
\newcommand{\lco}{L$_{CO}$}
\newcommand{\srcs}{$\sum \chi_r^2$}
\newcommand{\mum}{160~$\mu$m}
\newcommand{\htwo}{H$_2$}
\newcommand{\tmb}{T$_{\rm mb}$}
\shorttitle{High-resolution CO in the SMC}
\shortauthors{Muller et al.}

\begin{document}

%% LaTeX will automatically break titles if they run longer than
%% one line. However, you may use \\ to force a line break if
%% you desire.

\title{Characterizing the Low-Mass Molecular Component in the Northern
  Small Magellanic Cloud}

%% Use \author, \affil, and the \and command to format
%% author and affiliation information.
%% Note that \email has replaced the old \authoremail command
%% from AASTeX v4.0. You can use \email to mark an email address
%% anywhere in the paper, not just in the front matter.
%% As in the title, use \\ to force line breaks.
\author{
Muller, E.\altaffilmark{1,2}\footnote{Bolton Fellow, ATNF},  
Ott, J.\altaffilmark{3,4}\footnote{J\"urgen Ott is a Jansky Fellow
  of the National Radio Astronomy Observatory. The National Radio  Astronomy Observatory is a facility of the National Science
  Foundation operated under cooperative agreement by Associated
  Universities, Inc.}, 
Hughes, A.\altaffilmark{2,5}, 
Pineda, J.L. \altaffilmark{6}\footnote{NASA Postdoctoral Program Fellow},  
Wong, T.\altaffilmark{7,2,8}, 
Mizuno, N.,\altaffilmark{1,9}, 
Kawamura, A.,\altaffilmark{1}, 
Mizuno, Y.\altaffilmark{1},
Fukui, Y. \altaffilmark{1},
Onishi, T. \altaffilmark{1,10}
\and
Rubio, M. \altaffilmark{11}}
\affil{$^1$ Department of Astrophysics, Nagoya University, Furo-cho,  Chikusa-ku, Nagoya 464-8602, Japan}
\affil{$^2$Australia Telescope National Facility, CSIRO, P.O. Box 76, Epping, NSW, 1710, Australia}
\affil{$^3$National Radio Astronomy Observatory, P.O. Box O, 1003 Lopezville Road, Socorro, NM 87801, U.S.A.} 
\affil{$^4$ California Institute of Technology, 1200 E. California  Blvd, Caltech Astronomy, 105-24, Pasadena, CA, 91125, USA.} 
\affil{$^5$ Centre for Supercomputing and Astrophysics, Swinburne  University of Technology, Hawthorn, VIC 3122, Australia}
\affil{$^6$ Jet Propulsion Laboratory, California Institute of Technology, 4800 Oak Grove Drive, Pasadena, CA , USA} 
\affil{$^7$ School of Physics, University of New South Wales, Sydney, NSW, 2052, Australia}
\affil{$^8$ Department of Astronomy, University of Illinois, Urbana,  IL, 61801, USA.}
\affil{$^9$ National Astronomical Observatory of Japan, 2-21-1 Oswa, Mitaka, Tokyo, 181-8588, Japan}
\affil{$^{10}$ Department of Physical Science, Osaka Prefecture University, Gakuen 1-1, Sakai, Osaka 599-8531, Japan.}
\affil{$^{11}$ Departamento de Astronomia, Universidad de Chile, Casilla 36-D, Santiago, Chile}

%% Notice that each of these authors has alternate affiliations, which
%% are identified by the \altaffilmark after each name.  Specify alternate
%% affiliation information with \altaffiltext, with one command per each
%% affiliation.

%% Mark off your abstract in the ``abstract'' environment. In the manuscript
%% style, abstract will output a Received/Accepted line after the
%% title and affiliation information. No date will appear since the author
%% does not have this information. The dates will be filled in by the
%% editorial office after submission.

\begin{abstract}
We present here the first results from a high-resolution survey of the $^{12}$CO(J=1-0) emission  across the northern part of the poorly-enriched Small Magellanic Cloud, made with the ATNF Mopra telescope. Three molecular complexes detected in the lower resolution NANTEN survey are mapped with a beam FWHM of $\sim$42$''$, to sensitivities of approximately 210 mK per 0.9 \kms\ channel, resolving each complex into 4-7 small clouds of masses in the range of M$_{vir}\sim$10$^{3-4}$\msol and with radii no larger than 16 pc.  

The northern SMC CO clouds follow similar empirical relationships to the southern SMC population, yet they appear relatively under-luminous for their size, suggesting that the star-forming environment in the SMC is not homogeneous. Our data also suggests that the CO cloud population has little or no extended CO envelope on scales$\gtrsim$30 pc, further evidence that the weak CO component in the north SMC is being disassociated by penetrating UV radiation.

The new high-resolution data provide evidence for a variable correlation of the CO integrated brightness with integrated \hi\ and \mum\ emission; in particular CO is often, but not always, found coincident with peaks of \mum\ emission, verifying the need for matching-resolution \mum\ and \hi\ data for a complete assessment of the SMC H$_2$ mass.
%We identify molecular material within the body of the SMC that is distributed over two distinct and widely separate velocity ranges within closely spaced lines of sight. Such a bifurcation suggests the action of an energetic process sometime in the history of the SMC.
\end{abstract}

\keywords{ISM: molecules --- Magellanic Clouds --- galaxies: dwarf ---
  stars: evolution --- radio lines: galaxies}

\section{Introduction}

Molecular clouds are associated with the very earliest stages of star-formation and the environment external to the molecular cloud (e.g. enrichment levels, Interstellar medium (ISM) density variations, ambient magnetic and UV fields) must also play a significant role in the star-formation process. To understand the extent these and other parameters affect star-formation, we must examine the evolution of molecular clouds within a range of different environmental conditions.

The Magellanic Clouds are relatively metal poor; the LMC has a metallicity that is $\sim$30\% of the solar value and the SMC about 10\% of the solar value \citep[e.g.][]{larsen}.  Furthermore, The molecular cloud population in the Magellanic Clouds are unique in the extra galactic molecular cloud ensemble; \cite{bolatto08} show that the molecular clouds in the south-west of the SMC are  anomalously weak and under-luminous in comparison with other nearby extra galactic systems. At a distance of $\sim$60 kpc  \citep[e.g. ][who report  D$_{SMC}\sim$ 63 kpc, however we use D$_{SMC}\sim$ 60 kpc for consistency with existing liteurature]{cioni}, the SMC hosts a unique laboratory with an un-enriched ISM that is similar to that found in early-epoch and unevolved galaxies. 

The molecular (i.e. \htwo) component of galactic systems is most commonly estimated indirectly from observations of other tracers:  e.g. from FIR or mm/sub-mm datasets and related to the CO fraction via empirically-derived relationships. \cite{leroy07} recently combined \hi\ and \mum\ observations of the SMC to estimate the molecular distribution and the $X_{SMC}$ factor (where the \emph{$X$ factor} is an empirical conversion between CO and \htwo; $X=N(H_2)/W_{CO}$ cm$^{-2}$[K\kms]$^{-1}$). They find an  $X_{SMC}$ factor that is broadly in agreement with those derived from virial mass estimates, if slightly higher. A higher $X_{SMC}$ factor implies that the actual mass of the CO clouds is \emph{greater} equal to their measured virial mass, which may then require some magnetic support to prevent complete fragmentation. \cite{leroy07} (See also \cite{bot} and \cite{israel2003} - hereafter SKP-X,  to indicate paper X of the SEST key project) go on to suggest that the \htwo\ component, which provides more effective self-shielding than does CO, resides in larger and weaker extended envelopes and is not as effectively traced by strong CO emission. In this scenario, the lower dust-to-gas ratio allows $UV$ radiation to penetrate further into the ISM and disassociate the outer CO layers  to a level that is not detectable by current observations \citep[See also ][]{pak}.  

However \cite{pineda} made a study of the $X$ factor as a function of a strongly varying UV field within the LMC (nearby to the bright star-forming region, 30 Doradus) and found that the value of $X$ and has a mean value of 3.9$\times$10$^{20}$cm$^{-2}$[K\kms]$^{-1}$; $\sim$2 $X_{Gal}$ but does not vary with the strength of the UV field. Given the apparent lower metallicity of the SMC relative to the LMC \citep[e.g.][]{rolleston_smc,rolleston_lmc}, we expect a higher $X_{SMC}$ factor and indeed, $X_{SMC}$ estimates are of the order of 10$^{21}$ (e.g \citealt[][]{leroy07,rubio1991,israel1993}-hereafter SKP-I,), with some exceptions where $X_{SMC}$  is found to be $\sim$10$^{20}$ (e.g. SKP-X).

Generally however, the existing measurements of the global molecular component of the SMC have been, by virtue of the weak emission, confined to very specific regions, often comprising a single pointing \citep[e.g.][and references therein]{rubio1996,rubio2000,israel1993,israel2003}.  In particular \citeauthor{rubio1993a} (\citeyear{rubio1993a}; hereafter SKP-II), demonstrated the importance of high-resolution observations in resolving the molecular clouds and refining their basic morphological properties. 

The NANTEN telescope has mapped the largest area of the SMC at any molecular transition \citep{mizuno2001,blitz}, covering the SMC bar, and parts of the eastern side. This 4 m telescope has a resolution of $\sim$2.6' at the $^{12}$CO(J=1-0) line (sub-tending 44 pc at D$\sim$60 kpc), and so despite the excellent coverage, there remain some questions regarding the extent to which NANTEN can resolve molecular clouds in the SMC.

The existing targeted studies cannot form a resolved and comprehensive statistical study of the general molecular component in the SMC. It is therefore still impossible to completely understand the properties of the molecular clouds in the SMC as an ensemble, or to understand any large-scale systematic variations of the molecular cloud population throughout the SMC. It is with these issues in mind that we have undertaken to map the molecular component in the northern part of the SMC at high spatial resolution ($\sim$42\asec). This project forms the SMC branch of the molecular survey of the Magellanic System:  'Magellanic Mopra Assessment' \citep[MAGMA;][]{ott, pineda}.

In Section \ref{sec:2}, we detail the observations and data reduction. In Section \ref{sec:3} we discuss the methods used to process and analyze the resulting data cubes. In Section \ref{sec:4} we briefly report on the results. Section \ref{sec:5} contains a discussion of the detected CO population in the context of the SMC ISM, and Section 6 summarizes our findings.

\section{Observations and Data Reduction}\label{sec:2}
Observation targets were selected from clouds in the north of the SMC that showed an integrated CO intensity of at least 0.2 K\kms\ \citep[as measured by NANTEN][]{mizuno2001}. These clouds include those labeled NE-1 and NE-3; mapped at 8'.8 by \cite{rubio1991} and we further extend the ensemble to include a previously unlabeled cloud, which we call NE-4. NE-4 is visible in maps by NANTEN \citep{mizuno2001} but has no specific reference in that work. 

Observations were conducted during  2007, August and September with the
the Mopra 22 m Telescope \footnote{The Mopra radio telescope is part of the Australia Telescope which is funded by the Common-
wealth of Australia for operation as a National Facility managed by CSIRO. The University of New South Wales Digital filter bank used for the observations with the Mopra Telescope was provided with support from the Australian Research Council.} in Australia, and the  {\sc UNSW-mops} spectrometer\footnote{The University of New South Wales Digital Filter Bank used for the observations with the Mopra Telescope was provided with support from the Australian Research Council.}. Observations were made using the 'zoom' mode, where the 138 MHz-wide band is sampled with 4096 channels.  Pointing accuracy was maintained by frequent (every 1.5 hours) observations of the SiO maser Upsilon Mensa, so that errors were typically less than 5\arcsec. Calibration and measurements of the T$_{sys}$ were made from frequent ($\sim$15 minute interval) measurements of a hot load. The system temperature and data quality were monitored and calibrated using an in-horn diode calibrator.

Areas were scanned in overlapping 5\amin$\times$5\amin\ regions. Each area was observed twice with a sampling rate of 2 seconds (equivalent to three integrations per $\sim$ 30" telescope beam-width), with orthogonal scanning directions to reduce scanning artifacts. Observations of $^{12}$CO(1-0) in Orion KL were made once each observing session to monitor telescope stability. The resulting peak flux measurements were found to be self-consistent within 10\%, and also with observations made by the SEST telescope\footnote{http://www.apex-telescope.org/sest/html/telescope-calibration/calib-sources/orionkl.html} \citep[see also][]{ned}. The data were baseline-corrected with a 0th order polynomial fit and doppler-corrected using the {\sc aips++ livedata } package. 

\begin{figure}
\centerline{\resizebox{7cm}{!}{\includegraphics{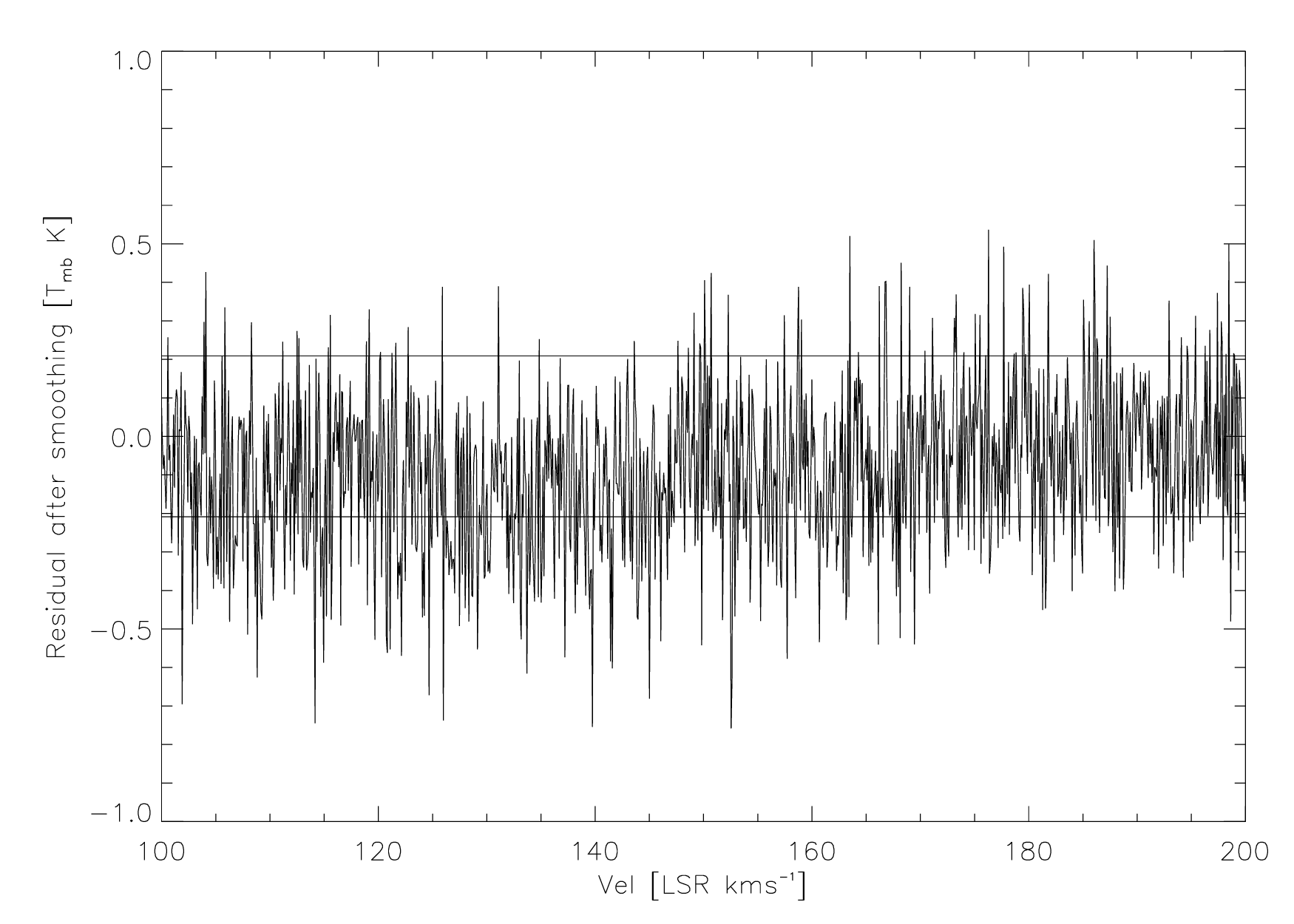}}}
\caption{Example of component that is \emph{removed} from a sightline, following the spatial frequency-domain smoothing process described in the text. This spectrum is towards NE-1a (00:59:45 -71:44:40;J2000). The two horizontal lines indicate $\pm$1 $\sigma$ levels. \label{fig:resids}}  
\end{figure}

The cubes were formed using median weighting with the {\sc gridzilla} package (beam-weighted gridding, and further smoothed with a radius-truncated Gaussian kernel, both with a FWHM of $\sim$35 arcseconds). Some of the spectra have a low-order and low-amplitude baseline variation present, which causes spurious detections in analysis software used later in this study. To remove this low-order baseline, the data cubes were differenced with a duplicate data cube, which was heavily smoothed in the frequency domain (width=26\kms). As the expected linewidths of the CO emission were of the order of a few \kms \citep[e.g.][]{rubio1996}, this step does not significantly affect the measurement and results in a relatively minor adjustment to the mean power per channel (which was always less than the one-sigma noise level, as shown in Figure~\ref{fig:resids}). Note also that any systematic or random errors arising from the baseline correction made here have not propagated through to the final brightness temperature estimates). Making this spectral smoothing-and-subtraction step improved the reliability of the cloud-searching algorithm by forcing the mean power outside channels containing emission to be approximately zero, and reduces false detections associated with a slowly-varying spectral baseline.

%Note that the smoothing step is made before subsequent velocity and position binning or smoothing. 

Finally, to improve sensitivity, the data cubes were binned to 0.9 \kms\ and convolved with 30\asec\ kernel, yielding an effective resolution of 42\asec\ ($\sim$12 pc). The data cube was formed with a pixel size of 15\asec\ ($\sim$4 pc). It should be noted that {\sc mops} uses a large digital filter bank, rather than a traditional auto-correlator that operates in the Fourier domain. As a result, the frequency resolution is well approximated by the frequency channel spacing. Assuming an main-beam efficiency of 0.47 \citep[e.g.][]{annie} at 115 GHz for the Mopra telescope, the resulting RMS level per channel is $\sim$ 210 mK (\tmb).

\begin{figure*}
\centerline{\resizebox{15cm}{!}{\includegraphics{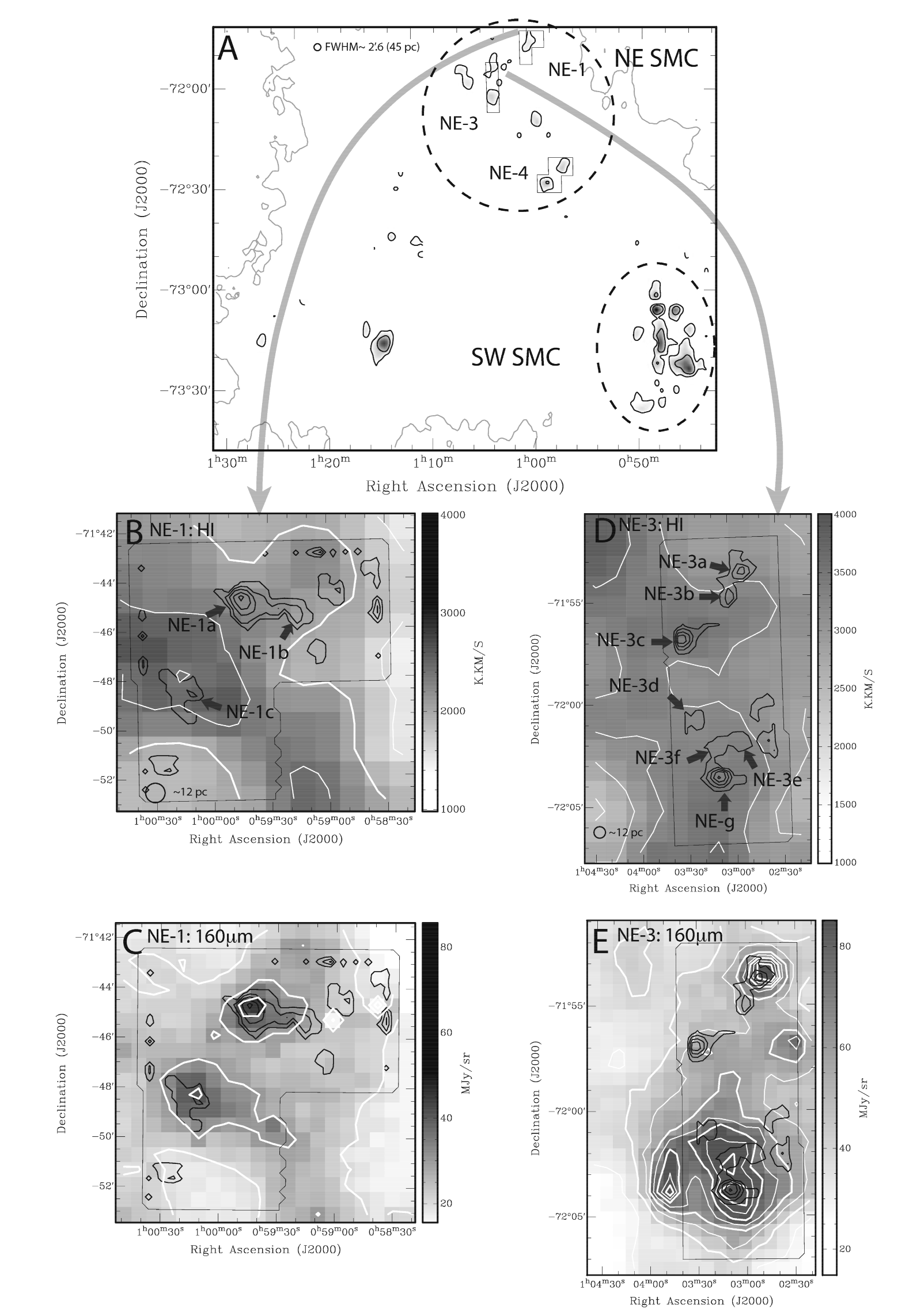}}}
\caption{\emph{Panel A}: map of $^{12}$CO(1-0), from \cite{mizuno2001} with   boxes locating the areas which were observed by Mopra. Black contours are  0.3, 0.9 K\kms. NE-1 and  NE-3 are shown in panels {\emph C} and {\emph D}, see Figure~\ref{fig:ne4} for NE-4. The dotted ellipses clarify the terminology for 'Northeast' and 'Southwest' regions, as used in this report. The grey outer contour shows the 1$\times$10$^{21}$ \hi\ integrated intensity level.\\
\emph{Panels B,C}: Contours of Mopra CO observations (\emph{Black}) made around NE-1, overlaid on integrated \hi\ (\emph{B}) \citep{stanimirovicphd} and \mum (\emph{C}) observations \citep{spitzer}. CO contours show integrated brightness temperature levels 0.5+1 K\kms; White \hi\ and \mum\ contours are to aid visualization, and are at intervals of 400 K\kms\ and 10 Mjy Str$^{-1}$ respectively.\\
\emph{Panels D,E}: Contours of Mopra CO observations made around NE-3, overlaid on integrated \hi\ (\emph{D}) and \mum\ (\emph{E}) observations. CO contours show the 0.5+1 K\kms; \hi\ and \mum\ contours are to aid visualization, and are at intervals of 400 K\kms\ and 10 Mjy Str$^{-1}$ respectively.
\label{fig:ne23}}
\end{figure*}

\begin{figure*}
\centerline{\resizebox{18cm}{!}{\includegraphics{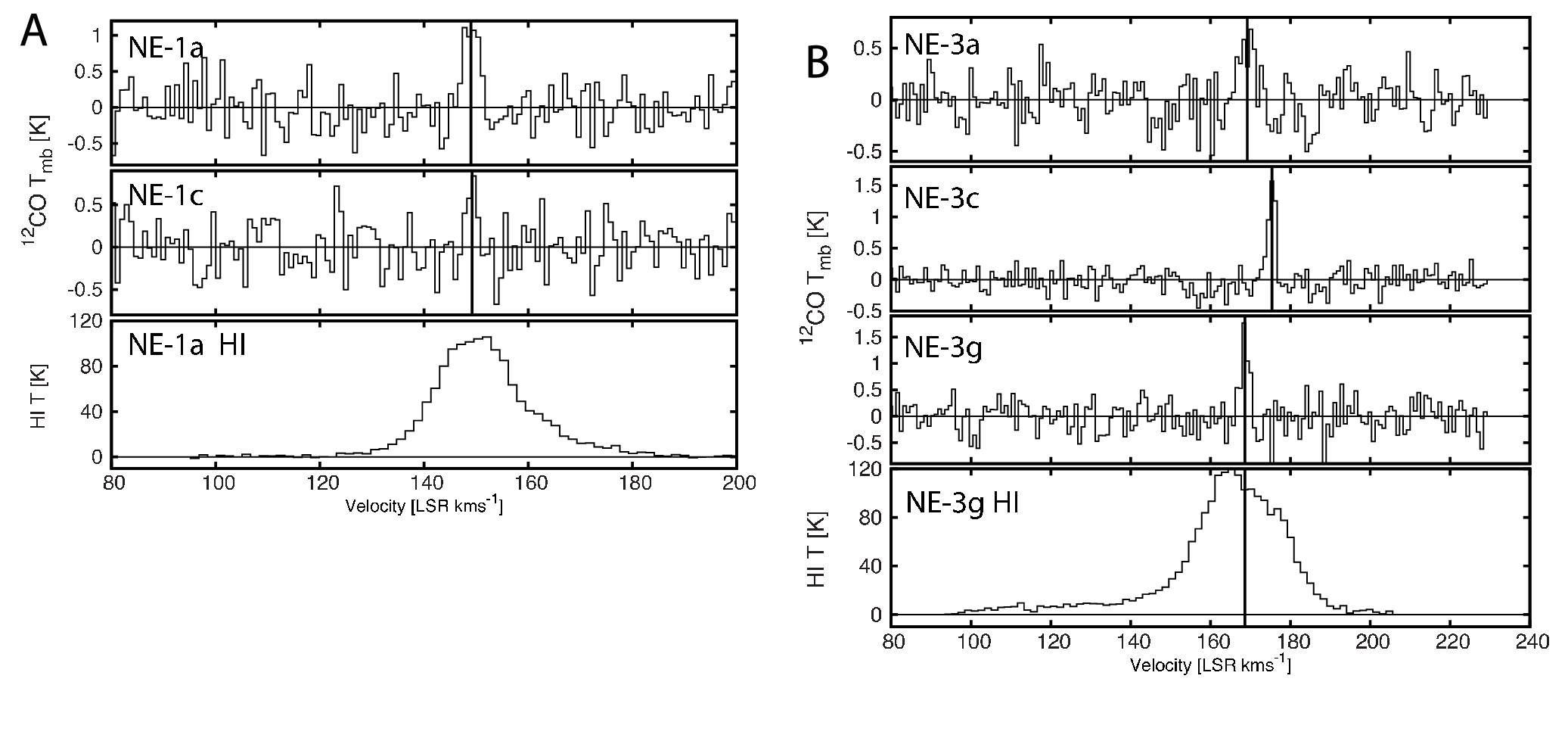}}}
\caption{
\emph{Panels A}: CO spectra towards peak positions of NE-1a and NE-1c, and \hi\ spectra towards NE-1a \emph{Bottom}. The velocity centroids are marked with a vertical line. Note that the \hi\ profile shown in the bottom is measured towards NE-1a and approximately the same throughout the entire of NE-1\\
\emph{Panels B}:  CO spectra towards peak positions of NE-3a, NE-3c and NE-3g, and \hi\ spectra towards NE-3g \emph{Bottom}. The CO velocity centroids are marked with a vertical line. As for \emph{Panel B}, the \hi\ profile is measured only towards NE-3g, but is approximately the same throughout the entire of NE-3.\\
\label{fig:ne23spect}}
\end{figure*}

\begin{figure*}
\centerline{\resizebox{18cm}{!}{\includegraphics{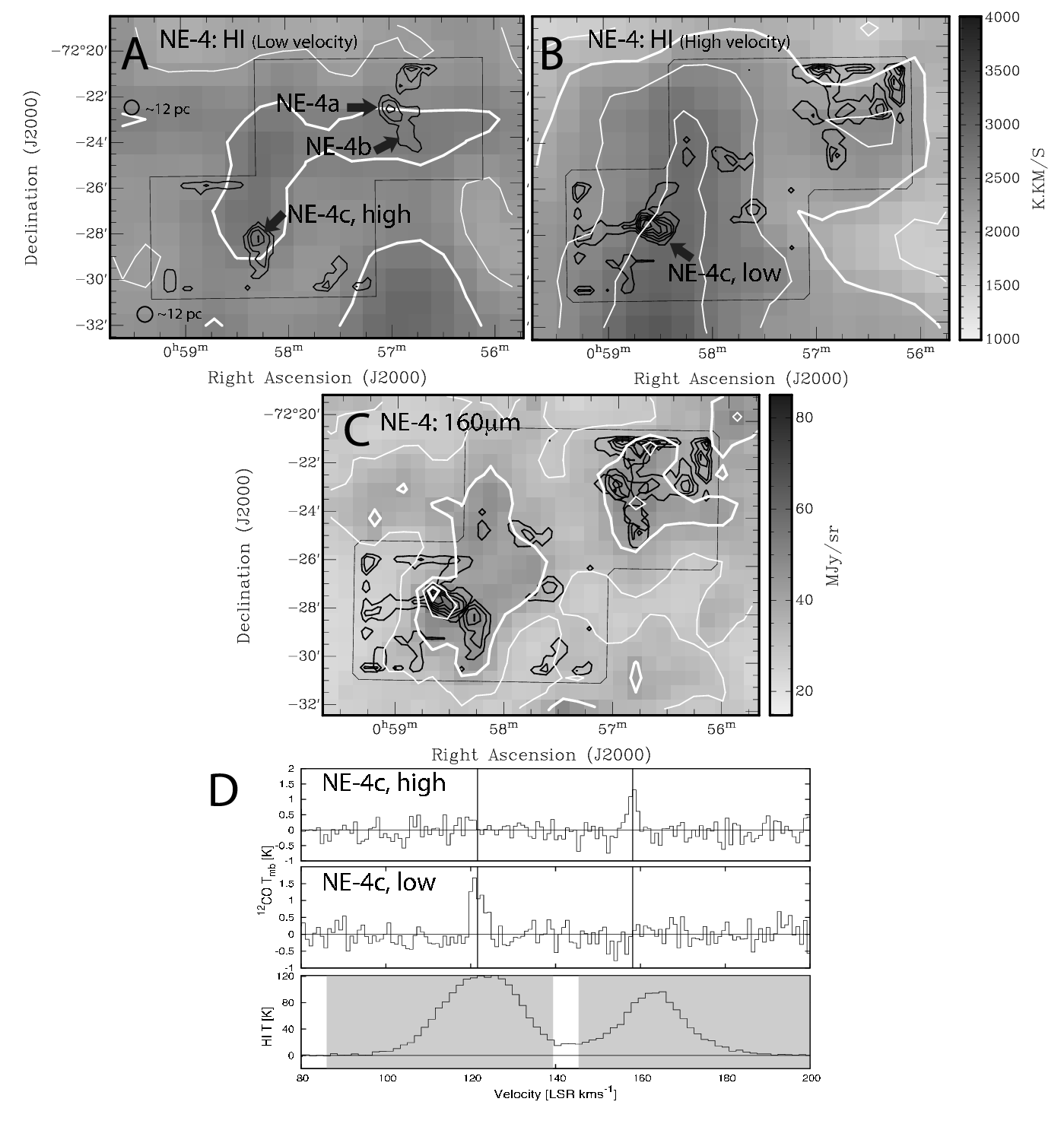}}}
\caption{ 
Contours of Mopra CO observations made around NE-4, overlaid as black contours (0.5+1 K\kms) on integrated \hi\ (\emph{A, B}) \citep{stanimirovicphd} and \mum\ (\emph{C}) observations \citep{spitzer}. The CO emission is broken into the high (\emph{Panel A}) and low (\emph{panel B}) velocity component and shown with the \hi\ also integrated over the high and low velocity ranges (shown as grayed regions in the top spectra of Panel \emph{D}) . Both \hi\ and CO components are shown together in \emph{Panel C}. White contours showing the \hi\ in intervals of 400 K\kms\ (in \emph{Panels A,B}), and in intervals of 10 Mjy str$^{-1}$ (in \emph{Panel C}) to aid visualization.\label{fig:ne4}} 
\end{figure*}

The maps and spectra for regions NE-1, NE-3 and NE-4 are shown in Figures~\ref{fig:ne23},~\ref{fig:ne23spect} and \ref{fig:ne4}. These figures show the integrated intensity for pixels that were detected by \cp\ as containing emission only (See Section~\ref{sec:3}).

\section{Cloud Identification Procedure}\label{sec:3}
The literature contains various methods of determining the radius of CO clouds detected in the SMC. \cite{rubio1991} and \cite{mizuno2001} choose to define the extent of the cloud based on the sensitivity of the data: pixels brighter than 3$\sigma$ in the integrated intensity maps are added and represent the area of the cloud. We attempt here an approach that is slightly less dependent on the quality of the observational data, where clusters of spectral lines containing emission are detected using a moment analysis by the \citep{cloudprops} analysis package \cp. We stress that \cp\ is used here only to objectively identify spectra containing emission within four adjacent velocity channels. \cp\ is not used for further analysis or interpretation and the observable parameters are measured from pixels that are identified by \cp\ as containing emission.

We apply \cp\ to identify regions containing signal greater than 1.5$\sigma$ (by setting the {\it thresh} parameter)  over at least 4 spectral channels and spread over an area equal to or greater than the beam size ($\sim42$\asec). We further examine cloud candidates by eye to eliminate spurious or marginal identifications. Typically, one or two putative 'clouds' in each of the three regions were falsely identified by \cp, and were rejected.

We estimate the cloud radii, \ico\ and \lco\ by summing the area of pixels containing emission that are close to the same local integrated intensity maxima (i.e. we group together all pixels that are closest to the same local maxima). We use the common definitions of these parameters: Radius, R = $(Area/\pi)^{0.5}$, virial mass; where \mvir = 190R$\Delta$V$^2$ (assuming a $1/r$ density profile throughout the cloud and where the velocity width ($\Delta$V) is measured at the position of maximum integrated brightness); and CO luminosity \lco, estimated from the product of the mean integrated emission and cloud area.  As the Mopra beam size and shape is not well understood and the SMC cloud radii are closely matched to the Mopra beam-size, we do not present the de-convolved radii of the clouds. We instead incorporate the de-convolved radius in the measurement errors by calculating the upper and lower errors separately: the lower error is equal to the de-convolved radius minus the measurement error of 2 pc. The errors reported for peak temperature are measurement errors only: we do not attempt to propagate the errors of the radius measurements and recalculate peak temperature measurement errors arising from beam dilution.

Line characteristics of spectra towards the position of peak integrated brightness (main beam brightness temperature, \tmb; velocity width,  $\Delta$V; velocity centroid, V$_{\rm 0}$) are estimated by fitting a single-component Gaussian (using the Marquardt-Levenberg algorithm). Given the low Signal-to-noise (S/N) ratio of these CO spectra and their simple, single-component profiles, it is not appropriate or beneficial to seek to decompose the emission into more than one spectral component. The errors for observables obtained in this way are estimated as asymptotic errors using statistical algorithms developed by \cite{landman}. 

It should be noted that the low S/N of these observations propagates to large errors in the fitted parameters, typically of around 50\% or even greater. As such, the resulting best-fit Gaussian profiles were occasionally  narrower than the instrumental resolution of the telescope (e.g. cloud NE-3d). Table~\ref{tab:1} lists the cloud parameters : Peak \tmb, V$_{\rm 0}$, $\Delta$V, radius R, \mvir, \ico.

\begin{deluxetable*}{cclllllll}
\tabletypesize{\scriptsize}
\tablecaption{Northern SMC CO-cloud parameters\label{tab:1}}
\tablewidth{0pt}
\tablehead{
Cloud/complex & RA& Dec&Peak T$_{MB}$& V$_o$ &$\Delta$V
&Radius\tablenotemark{a}&\mvir
&\lco $\times$10$^{3}$\\
& (J2000) & J(2000)& [K]&[\kms] &[\kms] &[pc]&[$\times$10$^3$\msol]&[K\kms\ pc$^{2}$]
}
\startdata
Complex NE-1    &&&&&&\\
NE-1a & 00:59:50&-71:44:40&1.2 $\pm$ 0.2&149.0 $\pm$ 0.2&3.4$\pm$0.6&16$^{+2}_{- 3}$&40$^{+10}_{-10}$&1.5$^{+0.5}_{-0.7}$\\
NE-1b & 00:59:20&-71:45:30&0.8 $\pm$ 0.3&149.2 $\pm$ 0.3&2.1$\pm$0.8& 9$^{+2}_{- 5}$&8$^{+6}_{-7}$&0.3$^{+0.2}_{-0.3}$\\
NE-1c & 01:00:10&-71:48:40&0.9 $\pm$ 0.4&150.0 $\pm$ 0.3&2.0$\pm$0.7&13$^{+2}_{- 3}$&10$^{+7}_{-8}$&0.6$^{+0.2}_{-0.3}$\\\tableline\\
Complex NE-3/N76    &&&&&&\\
NE-3a & 01:02:50&-71:53:30&0.6 $\pm$ 0.1&169.2 $\pm$ 0.5&5$\pm$1&12$^{+2}_{- 3}$&60$^{+30}_{-30}$&0.6$^{+0.3}_{-0.4}$\\
NE-3b & 01:03:00&-71:55:00&0.7 $\pm$ 0.2&171.9 $\pm$ 0.3&2.4$\pm$0.6& 9$^{+2}_{- 4}$&11$^{+6}_{-8}$&0.3$^{+0.2}_{-0.3}$\\
NE-3c & 01:03:30&-71:57:00&1.7 $\pm$ 0.2&175.4 $\pm$ 0.1&2.1$\pm$0.3&11$^{+2}_{- 4}$&10$^{+3}_{-4}$&0.7$^{+0.3}_{-0.5}$\\
NE-3d & 01:03:20&-72:01:00&0.6 $\pm$ 0.2&168.6 $\pm$ 0.2&2.3$\pm$0.8& 7$^{+2}_{- 7}$&8$^{+6}_{-8}$&0.2$^{+0.1}_{-0.2}$\\
NE-3e & 01:03:00&-72:02:00&0.7 $\pm$ 0.2&168.3 $\pm$ 0.4&2.4$\pm$0.6& 9$^{+2}_{- 4}$&11$^{+6}_{-8}$&0.3$^{+0.2}_{-0.3}$\\
NE-3f & 01:03:10&-72:02:00&0.8 $\pm$ 0.2&168.1 $\pm$ 0.2&2.4$\pm$0.6& 9$^{+2}_{- 4}$&11$^{+6}_{-8}$&0.3$^{+0.2}_{-0.3}$\\
NE-3g & 01:03:10&-72:03:50&1.6 $\pm$ 0.3&168.6 $\pm$ 0.2&2.3$\pm$0.5&13$^{+2}_{- 3}$&14$^{+6}_{-7}$&1.0$^{+0.3}_{-0.5}$\\\tableline\\
Complex NE-4    &&&&&&\\
NE-4a$_{hi}$ & 00:57:00&-72:22:40&1.5 $\pm$ 0.3&153.9 $\pm$ 0.2&2.2$\pm$0.5&11$^{+2}_{- 4}$&11$^{+5}_{-6}$&0.6$^{+0.2}_{-0.4}$\\
NE-4b$_{hi}$ & 00:56:50&-72:24:20&0.8 $\pm$ 0.2&154.1 $\pm$ 0.3&1.8$\pm$0.6& 9$^{+2}_{- 5}$&6$^{+4}_{-5}$&0.2$^{+0.1}_{-0.2}$\\
NE-4c$_{hi}$ & 00:58:20&-72:28:10&1.3 $\pm$ 0.2&158.2 $\pm$ 0.2&2.3$\pm$0.5&15$^{+2}_{- 3}$&16$^{+7}_{-8}$&0.9$^{+0.3}_{-0.5}$\\
NE-4c$_{low}$ & 00:58:40&-72:27:40&1.6 $\pm$ 0.3&122 $\pm$ 1&5$\pm$2&14$^{+2}_{- 3}$&70$^{+60}_{-60}$&1.7$^{+0.5}_{-0.9}$\\
\enddata
\tablenotetext{a}{Lower errors for cloud radii include potential impact smearing by the Mopra beam (FHWM$\sim$12 pc)}

\enddata
\end{deluxetable*}
To provide some assurance of the effectiveness of our data preparation and analysis technique, we apply it also to data of the south-west SMC, taken with the SEST telescope \citep[][hereafter, SKP-III]{rubio1993b}. The analysis method we apply to the SEST dataset is almost identical to that used on the data presented here, with the exception of the smoothing-and-subtraction baseline correction step. This step was omitted primarily because the frequency-switching mode employed for the SEST observations generates a "negative" spectral artifact nearby to the real emission spectrum, and also because insufficient line-free channels exist to create a usable smoothing function (as the emission line occurs near the edge of the band). Instead, we have removed a second-order baseline, interpolated from line-free channels across velocities channels where emission occurs. 

We provide the results of this comparison in Table~\ref{tab:comparison} and show that our method produces values that are consistent to within approximately 10\% with the reported in SKP-III. Note that errors are not quoted in SKP-II and SKP-III, however the RMS for the SKP observations is cited as 200 mK, and later papers in that series cite \tmb\ accuracies of 15-20\%. 
\begin{deluxetable}{lll}
\tabletypesize{\scriptsize}
\tablecaption{Confirmation of cloud analysis algorithm\label{tab:comparison}}
%\tablewidth{0pt}
\tablehead{
Cloud/Object & This paper& SKP-II, SKP-III$^{a}$ \\
}
\startdata
LIRS36	&	&\\
\tmb\ [K]				&	2.4$\pm$0.1	&	2.37\\
$\Delta$V [K\kms]			&	3.1$\pm$0.1	&	3.1\\
R [pc]					&	16.3$\pm$0.2	&	18.6\\
\lco [$\times$ 10$^3$ K\kms\ pc$^2$]	&	2.0$\pm$0.5	&	2.8\\\tableline\\

LIRS49	&	&\\
\tmb\ [K]				&	1.6$\pm$0.1	&	1.88\\
$\Delta$V [K\kms]				&	4.8$\pm$0.1	&	4.8\\
R [pc]					&	17.3$\pm$0.2	&	18.6\\
\lco [$\times$10$^3$ K\kms\ pc$^2$]		&	6$\pm$1	&	6.41\\\tableline\\

SMCB1	&	&\\
\tmb\ [K]				&	1.4$\pm$0.1	&	1.38\\
$\Delta$V [K\kms]				&	2.75$\pm$0.1	&	2.88\\
R [pc]					&	15.2$\pm$0.2	&	13.8\\
\lco [$\times$10$^3$ K\kms\ pc$^2$]		&	1.3$\pm$0.4	&	1.21\\\tableline\\

N88 &	&\\	s
\tmb\ [K]				&	0.79$\pm$0.1	&	0.77\\
$\Delta$V [K\kms]				&	1.63$\pm$0.2	&	1.6\\
R [pc]					&	Unresolved.		&	Unresolved.\\
\lco [$\times$10$^3$ K\kms pc$^2$]		&	0.5$\pm$0.2	&	0.54\\

\enddata
\tablenotetext{a}{Note that measurement errors are not quoted in SKP-II or SKP-III, however the 1$\sigma$ noise temperature of those observations is $\sim$ 0.2 K. Note that other papers in that series (SKP-1, SKP-X) suggest a 15\%-20\% error in flux calibration. Linewidth ($\Delta$V) errors for the objects above are only cited in SKP-V, which are different, but similar to the $\Delta$V values reported in SKP-II and SKP-III. These errors quoted in SKP-V are typically $\sim$0.2 \kms.}
\end{deluxetable}

\section{Properties of CO clouds in the SMC}\label{sec:4}
Table~\ref{tab:1} shows that the detected and most massive clouds in the northern SMC are relatively small, with virial masses in the 10$^{3-4}$\msol\ range. This is at the extreme low end of an extra-galactic sample examined by \cite{bolatto08}, where the virial masses of the extra-galactic molecular cloud ensemble  exist over a range of $\sim$10$^{4-6}$\msol. We see then that the molecular cloud population in the northern SMC can be regarded as an extreme population relative to the south-west SMC, and more so in the extra-galactic context.

Figures~\ref{fig:ne23},~\ref{fig:ne23spect} and \ref{fig:ne4} show the relative distributions and $^{12}$CO(J=1-0) emission spectra of the three regions NE-1, NE-3 and NE-4. Also shown are the velocity-integrated I$_{CO}$ maps overlaid on \mum\ and \hi\ data, as tracers of the big-grain (BG) dust abundance and temperature, and the ubiquitous ISM, respectively. Note that the \hi\ maps are of much poorer resolution ($\sim$98") than the CO dataset and are used here to aid a contextual discussion of the relationships between \hi, CO and \mum\ tracers. The \mum\  data is obtained directly from the S$^3$MC archive \citep{spitzer}\footnote{http://celestial.berkeley.edu/spitzer/} and have not been rigorously processed to remove, for example, the weak (but very complicated) contribution from foreground cirrus \citep[See][]{leroy07}.

\subsection{NE-1} 
We see in Figure~\ref{fig:ne23} (\emph{Panels B} and \emph{C}) the spatial relationship of NE-1 with \hi\ and \mum\ respectively. The CO emission from this region is relatively weak, with a peak \tmb\ of barely over 1 K. The detected CO clouds are spatially coincident with the strongest sources of \mum\ emission in this region (Figure~\ref{fig:ne23}B), although there are some significant departures from a perfect correlation, most likely due to varying dust temperature and/or small-scale fluctuations in  \hi\ column density that is not obvious from Figure~\ref{fig:ne23}C. Even so, and bearing in mind that the spatial resolution of the \hi\ observations is 98"; approximately three times that of the CO observations, Figure~\ref{fig:ne23}C shows the \hi\ integrated brightness varies only very slowly across the region with little evidence of any significant variations in intensity towards the CO sources.

We estimate the virial masses of the clouds (NE-1a, NE-1b and NE-1c) to be 4.0$\pm$1$\times$10$^4$\msol, 8$^{+6}_{-7}\times$10$^3$\msol\ and 10$^{+7}_{-8}\times$10$^3$\msol. Summed together, the mass of this complex measured by Mopra is approximately a factor of 50 below the virial mass estimated by \cite{rubio1991}. The combination of a lower resolved velocity dispersion (by a factor of  $\sim$3) and lower radius (by a factor of $\sim$6) will account for the disagreements in estimated virial masses, however we explore the wider-scales of this complex further in Section~\ref{sec:lowresdata}.

The emission profiles associated with NE-1a and NE-1c are shown in Figure~\ref{fig:ne23spect} (\emph{Panels A,B}) and these observations have enabled the velocity components of the emission regions to be separated into much narrower profiles than was previously known. We see that the CO peaks generally coincide with \hi\ peaks, although not always at exactly the peak velocity.
\subsection{NE-3}
The CO emission from this region is shown overlaid on \hi\ integrated intensity and on \mum\ brightness in Figure~\ref{fig:ne23} (\emph{Panels D,E} respectively). The Figures show a slightly larger area than was mapped by Mopra to include the full extent of an apparently coherent \mum\ emission region. We resolve up to seven peaks of emission distributed throughout this region. 

We find a total virial mass of 1.25$^{+6.3}_{-7.3}\times$10$^5$\msol; a factor of $\sim$100 less than that reported by \cite{rubio1991}. The difference in measured velocity widths and radii cannot account for the difference in the estimated virial masses, however, there appears to be some differences in the CO maps from  \cite{mizuno2001} and \cite{rubio1991}. After allowing for different beam sizes we find that the morphology of the NANTEN results are more consistent with the high resolution results presented here. We note that the NANTEN maps show additional CO clouds that seem confused or absent in the \cite{rubio1991} study, and are approximately 10' east of the area mapped with Mopra for the present study.    We present a study of the larger-scale distribution of CO in the context of the NANTEN dataset in Section~\ref{sec:lowresdata}. 

%It is worth noting that the luminosity of NE-3 as measured by \cite{rubio1991}: 3.24$\times$10$^{4}$ K  \kms, with a radius of 155 pc, corresponds to a mean integrated CO brightness of 0.43 \kkms over the $\sim$75x10$^3$pc$^2$ area of the cloud. This brightness level is detectable by these Mopra data after binning over the appropriate velocity range (162-182 \kms) to a achieve sensitivity of $\sim$40mK, but no such extended envelope is identified.. We discuss the subject of missing CO later in section~\ref{sec:5} and we anticipate that additional, deeper and wider-field observations by Mopra will reveal the true extent of CO emission from NE-3. 

The distribution of the \mum\ emission associated with NE-3 again shows a generally good association of the peak \mum\ and CO (Figure~\ref{fig:ne23}E), and there are again some notable deviations from a good correlation. The \hi\ data  (Figure~\ref{fig:ne23}D) shows a rather slowly-varying and uncorrelated distribution but are not of sufficient resolution to obtain a good estimate of the relationship of CO and H$_{2}$ on these scales observed with Mopra.

\subsection{NE-4}
The measured CO emission from NE-4 is shown in the top of Figure~\ref{fig:ne4} as contours overlaid on \hi\ integrated over 85$<$V$_{LSR}<$140 (\emph{Panel A}) and 145$<$V$_{LSR}<$200 (\emph{Panel B}). All CO integrated over the line of sight is shown overlaid on \mum\ emission on \emph{Panel C} \citep{spitzer}.

Studies by \cite{mizuno2001} resolve this complex into two clouds, although this work did not discuss this region in detail. We show here in Figure~\ref{fig:ne4} that this region is now resolved further into four clouds, including a pair that shows evidence for easily-separable high and low velocity components, as shown in Figure~\ref{fig:ne4}D.

\emph{Panel D} of Figure~\ref{fig:ne4} shows that the CO emission peaks correspond closely, although not exactly in velocity to peaks of \hi. This suggests that the molecular and neutral components are to some extent, co-moving with the \hi. The projected spatial separation of the high and low velocity components of the two CO clouds is $\sim$30 pc. The fact that two separate regions are detected in close proximity may be evidence for a recent energetic event which has bifurcated a formed molecular cloud, however we will defer a more dedicated analysis of the velocity structure of the ISM at this location to a later paper.

The distribution of the \mum\ emission shown in Figure~\ref{fig:ne4}C, shows local maxima approximately spatially coincident with the CO emission, however once again, the resolution of the \hi\ data  (Figure~\ref{fig:ne4}C)  is too poor to permit a reliable estimate of the correlation of H$_2$ and CO on the $\sim$30 pc scales of the CO data.  %Of interest is the fact that the morphology of the \mum\ distribution appears more consistent with \hi\ of the \emph{higher} velocity range (Figure~\ref{fig:ne4}C), suggesting a stronger correlation with this velocity component \citep[see also][]{leroy07}.  

\section{Analysis and discussion}\label{sec:5}
We now have a means to examine the variations of the molecular cloud population throughout two widely separated regions in the SMC. Its dynamic and perturbed evolution suggests that different parts of the SMC may have evolved under slightly different conditions, and these new results enable a new level of analysis into the ISM of the SMC, specifically in the context of star-formation in its low-metallicity environment. The compact nature of the detected CO population contrasts the relatively more extended dust emission and provides further impetus for future studies of the relationship of  CO, \hi\ and \htwo. This is particularly true in the case of the low-extinction ISM of the SMC, where extended CO is more easily photo disassociated \citep[e.g. see][]{maloney,mckee,pak}. However we are now able to make some general observations regarding the distribution of small-scale CO and \mum\ brightness - in particular reference to the $X_{SMC}$ factor, and we are now in a position to examine the relationship of \hi\ and CO on $\sim$ 30 pc scales in the SMC.

\subsection{Regional variation of molecular cloud properties}\label{sec:lowresdata} 
\begin{figure*}
\centerline{\resizebox{18cm}{!}{\includegraphics{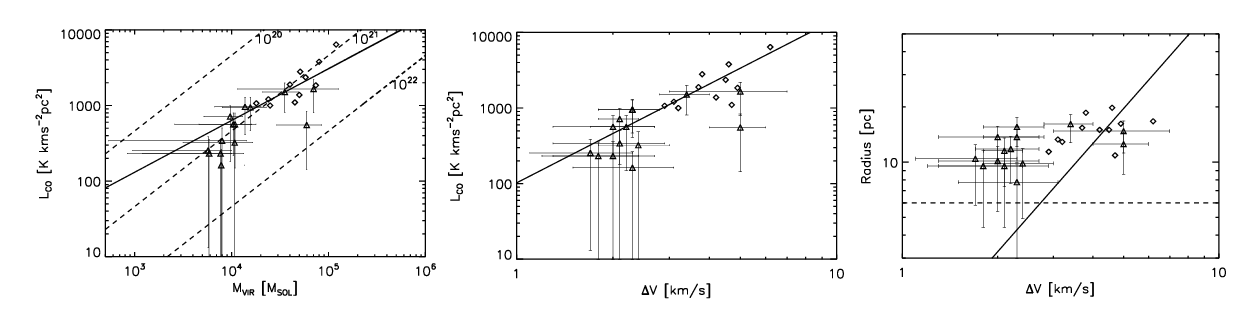}}}
\caption{Comparisons of cloud properties measured from this study (\emph{Triangles, with error-bars}), and those from the west of the SMC (SKP-III) (\emph{Diamonds}; no errors cited, see Table~\ref{tab:comparison} for estimations of errors of SKP-III data). (\emph{Left}): Virial mass and luminosity, along with dotted lines of constant X$_{co}$=10$^{20}$, 10$^{21}$ and 10$^{22}$ cm$^{-2}$ [\kkms]$^{-1}$. (\emph{Middle}): Linewidth and luminosity, and (\emph{Right}): Linewidth and size, where  the HPHW size of the Mopra beam ($\sim$ 6pc) is shown as a dotted line.
The solid straight lines are empirically-derived lines in SKP-III (no errors cited in the derivation of the empirical relationship).
 %We see that in the cases of the luminosity against virial mass (\emph{Left}) and luminosity against velocity dispersion (\emph{Center}), that generally the clouds measured in this study follow the same physical relationships to those in the south west of the SMC, derived in SKP-III. However, the clouds of this study are generally significantly narrower, suggesting a cooler evolutionary environment. The plot showing velocity dispersion against Radius (\emph{Right}) highlights that the clouds of this study do not conform as clearly to the same size-line with relation as those from the west of the SMC, and suggest that the true radii of the northern population may be much smaller than can be resolved with the Mopra telescope beam. 
\label{fig:plots}}
\end{figure*}

In Figure~\ref{fig:plots} we compare the characteristics of the CO clouds measured in this study to those in the South-West SMC, made using the SEST telescope in SKP-III. It should be noted that these observations by the SEST key projects are of comparable resolution and sensitivity to our Mopra observations of the north-west clouds. Overlaid on each of the panels in the figure are empirical relationships for the plotted properties as derived in SKP-III (log $\Delta$V=0.51 log R+0.04, log L$_{CO}$=2.17 log $\Delta$V+2.01 and log M$_{vir}$=1.46 log L$_{CO}$-0.09 ). Although the error bars for our data are large, we find that the ratios of luminosities and virial masses of the northern clouds (Triangles in Figure~\ref{fig:plots} left) are broadly consistent with those of the south-western population (plotted as diamonds). It is interesting that despite the substantial scatter in the  points, they appear to cluster more coherently around a constant X$_{CO}\approx$10$^{21}$ than the empirical fit derived from the southern population.
However we see that although Figure~\ref{fig:plots} (middle) shows a robust compliance to the empirical linewidth-luminosity relationship, the size-linewidth relationships (Figure~\ref{fig:plots}, right) do not, and this implies the clouds in the north SMC are significantly relatively under-luminous compared to their south-western counterparts. 

Other quantitative indicators of differences in the northern and southern star-formation environments exist; \cite{davies} compiled a catalog of \hii\ regions in the SMC that was later expanded by \cite{bica2008}.  A more thorough study of the relationship of the northern and southern \hii\ regions is deferred to a later study, however we conduct a Kolmogorov-Smirnov (K-S) test on the size of the \hii\ regions in the northern and southern parts of the SMC (i.e. northern clouds are those that are west of 1.1$^{hr}$  and north of $-$72\dg 45$'$, while  southern clouds are those west of 1.1$^{hr}$  and south of $-$72\dg 30$'$ - approximately the southern limit of the clouds examined by MAGMA-SMC) to test their morphological consistency. The KS test shows that the northern population has a probability of $\sim$3\% of being representative of the southern population and we find the \hii\ regions in the north are, on average, 60\% larger than \hii\ regions in the south, where the number populations of the north and south \hii\ datasets are 82 and 139, respectively. Generally, the sizes of the \hii\ regions can be modulated by local density of pressure inhomogeneities, or are even just a result of differences in the mean ionizing luminosities and this is a further indicator that the two populations are evolving within different ambient conditions.

\subsection{The compact nature of CO in the northern SMC}\label{sec:lowresdata}
To examine the compactness of the detected CO emission regions we compare CO luminosities from the lower-resolution and wider-field NANTEN maps to the results presented here. Our intention is to check that Mopra has detected the same total flux as NANTEN over the same scale ranges, which will reveal the quantity of extended and diffuse  CO. We are severely limited by the size of the field sampled by Mopra; the 5'$\times$5' arc-minute fields that have been measured in the Mopra survey subtend only two beams, or 2.5 pixels in the NANTEN dataset.

 Essentially, we are using  the NANTEN dataset as a control to check and compare the flux detected by the combination of the Mopra observations and \cp\ processing. We focus this test on the emission regions measured by Mopra and \cp\ \emph{only}; we form an I$_{CO}$ map using only spectra that have been identified by \cp\ as containing emission, thereby ignoring the noise-dominated parts of the map, which are instead set to zero, and then smooth the Mopra dataset to the same resolution as NANTEN (2'.6). The NANTEN maps extend over a much wider field than the Mopra maps, and we further limit the size of the field used in the following test to radii $\lesim$180 pc (approximately 4 beams in the NANTEN dataset, and 12 beams in the Mopra dataset) to avoid confusion by other nearby CO complexes that are far outside the Mopra complexes, but that will contaminate the NANTEN dataset through its wider beam.

We integrate the smoothed Mopra dataset and the NANTEN dataset over areas with incrementally increasing radii, focusing on each the complexes NE-1a,b,c, NE-3a,b,c,d,e,f,g and NE-4a,b,c in their entirety: e.g. we simply treat the NE-4 complex as a whole and increase the scale of the test beyond the dimensions of the complex. Examining the whole of the complexes in this way ensures that we are considering entire complexes as observed by NANTEN (i.e. all the components of NE-1 are confused by the NANTEN beam).  In general, we choose a point that is approximately central to each of the fields as our zero radius, although the exact position of the zero radius is generally unimportant.

\begin{figure*}
\centerline{\resizebox{18cm}{!}{\includegraphics{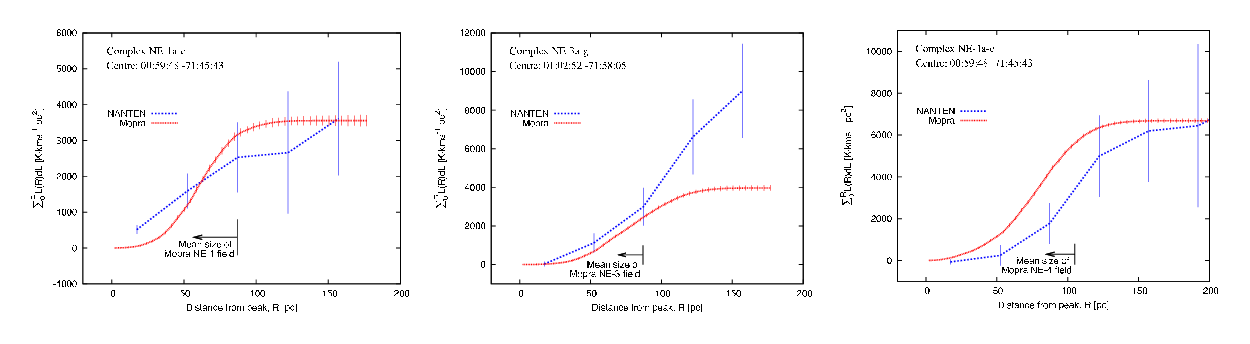}}}
 \caption{Azimuthally-integrated brightness plots, of NANTEN and (2'.6-smoothed) Mopra observations of complexes NE-1 (\emph {Left}), NE-3 (\emph {Middle}) and NE-4.(\emph {Right}). Error bars indicate the cumulative sum of the 1$\sigma$ noise level per pixel. Total power measured by Mopra for complexes NE-1 and NE-4 equate with that of NANTEN after some scale, indicating that the same total power is measured over the total area (despite the larger beam size of NANTEN), arguing for an absence of any larger extended CO envelope. Complex NE-3 shows extended emission measured by NANTEN from additional CO components outside and to the east of the area observed by Mopra. Note that the geometry of the observed regions are generally highly irregular, the indicated radii is that of an equivalent-area circle and not the maximum extents of the test areas. }
\label{azplots}
\end{figure*}

We show in Figure~\ref{azplots} the $L_{CO}(R)d$r for NE-1, NE-3 and NE-4 from the NANTEN and the smoothed Mopra data. We find that the power measured by Mopra becomes equal to that measured by NANTEN in complexes NE-1 and NE-4 within the sampled scale ranges (Figure~\ref{azplots}, left and right). The exact scale of the agreement is somewhat dependent on the distribution of the CO within the complex, but in both of these cases we find that Mopra measures the same total power as NANTEN within the $\sim$10' fields. %Note that as the position of the zero-radius is close to the middle of the entire complex, and is generally distant from the positions of each of the CO clumps in each complex, the  total integrated luminosity for each complex will not necessarily be equal to the sum of the individual clump luminosities.  

This result suggests that these the clouds (NE-1, NE-4) do not posses an extended CO envelope on radii much larger than those sampled in Figure~\ref{azplots} and listed in Table~\ref{tab:1}. The spatial range of this part of the study is limited to $\sim$10 arc minutes (where 10' subtends $\sim$170 pc at the $\sim$60 kpc distance of SMC), and although these high resolution observations do not completely  eliminate the possibility of a weak and extended envelope, the important result is that the total flux measured on the small scales by  Mopra is the same as that measured with the lower-resolution and wider-field NANTEN telescope. 
%As a result of the limiting sensitivity of NANTEN, (where the diffuse CO envelope fills the 2'.6 beam), low-level power that has a surface brightness less than 0.7$\times$10$^3$K\kms pc$^{-2}$ will be insufficiently bright to distinguish from the noise. This suggests that the clouds have, at the very most, core/envelope surface brightness ratios ranging between 2 - 20. We emphasize that this contrast change occurs over $\sim$ 6pc (i.e. approximately the radius of the clouds).

The azimuthally-integrated L$_{CO}$ of NE-3, on the other hand, (middle panel of Figure ~\ref{azplots}) shows a great difference in the fluxes measured by Mopra and NANTEN. The figure shows a strong divergence of the two datasets at the approximate mean size of the Mopra maps, which suggests that the  NANTEN data is being affected by a more extended envelope by a region to the east of NE-3 (see Figure~\ref{fig:ne23}). As much of this area is outside that observed by Mopra, it does not contribute to the Mopra dataset. Until this region is fully sampled, we are not able to make a conclusion regarding the existence of any extended CO envelope that may be associated with NE-3.

\subsection{CO and HI}\label{sec:co_hi}
\cite{wong} have made a thorough comparison of the properties of the \hi\ and CO emission profiles throughout the LMC. Most notably, they found that CO is detected only towards relatively high \hi\ column densities, where N(\hi)$\geq$10$^{21}$cm$^{-1}$, and that these column densities are a necessary, but insufficient requirement for the formation of CO in the LMC. The \hi\ found in the LMC exists largely in single-component profiles where it is appropriate to assume that where the centroids of the CO and \hi\ profiles coincide for a given line of sight, the two components are co-moving and co-evolving. In contrast, the neutral hydrogen component of the SMC is often complex in both position \emph{and} in velocity: the \hi\ profiles in the line of sight of the CO emission regions often contain a number of Gaussian components (See Figures~\ref{fig:ne23spect} and \ref{fig:ne4}), and it is inaccurate to assume that all the \hi\ found along a particular sight is necessarily associated with the detected CO emission. 

To isolate the \hi\ component that is associated with the CO, we \emph{assume} that an \hi\ and CO component must be co-moving, and constrain a multiple-component Gaussian fit with the CO emission velocity centroid as a fixed parameter. We do not assume that the CO and \hi\ components are similarly turbulence-broadened (i.e. $\Delta$V$_{CO}\neq\Delta$V$_{\sc hi}$).

For this analysis, we first smooth and re-grid the CO dataset to the same geometry as the \hi\ dataset (i.e. to a 98" beam, and 60" per pixel, which results in some dependence of adjacent pixels). As smoothing dilutes the peak brightness somewhat, we make this test only on the 8 brightest CO clouds; NE-1a,c, NE-3a,c,g and NE-4a,c$_{lo/hi}$.   

\begin{figure}
\centerline{\resizebox{10cm}{!}{\includegraphics{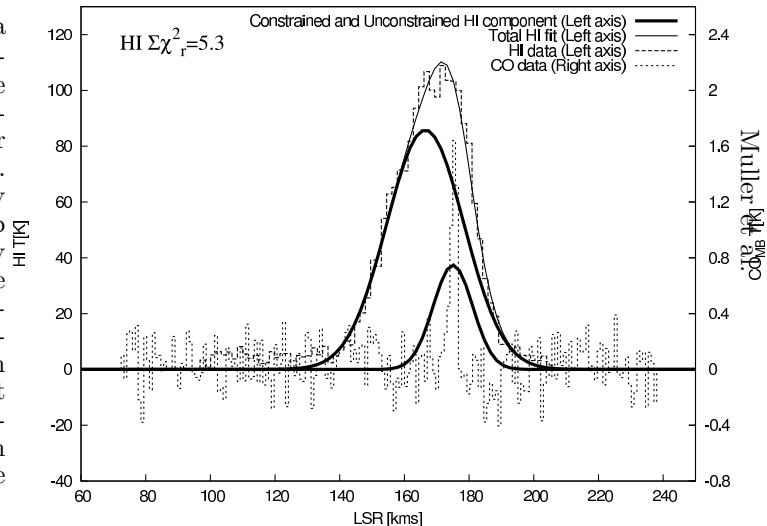}}}
\caption{Two-component fitting made to \hi, where one component is constrained by the velocity centroid (marked with a short vertical line). The \srcs\ parameter is shown, indicating a reasonable fit by constrained, two-component fit. The \hi\ emission over 80-130 km/s is real, but is not fitted nor included in the calculation of \srcs\ (see text for details). The heavy solid lines indicate the two fitted components. The lighter solid line is the total fitted result. The heavier dashed line shows the HI brightness temperature (left axis) and the lighter, noisier dashed line shows the CO brightness temperature
(right axis).
\label{fig:examplefit}}
\end{figure}

\begin{figure}
\centerline{\resizebox{8cm}{!}{\includegraphics{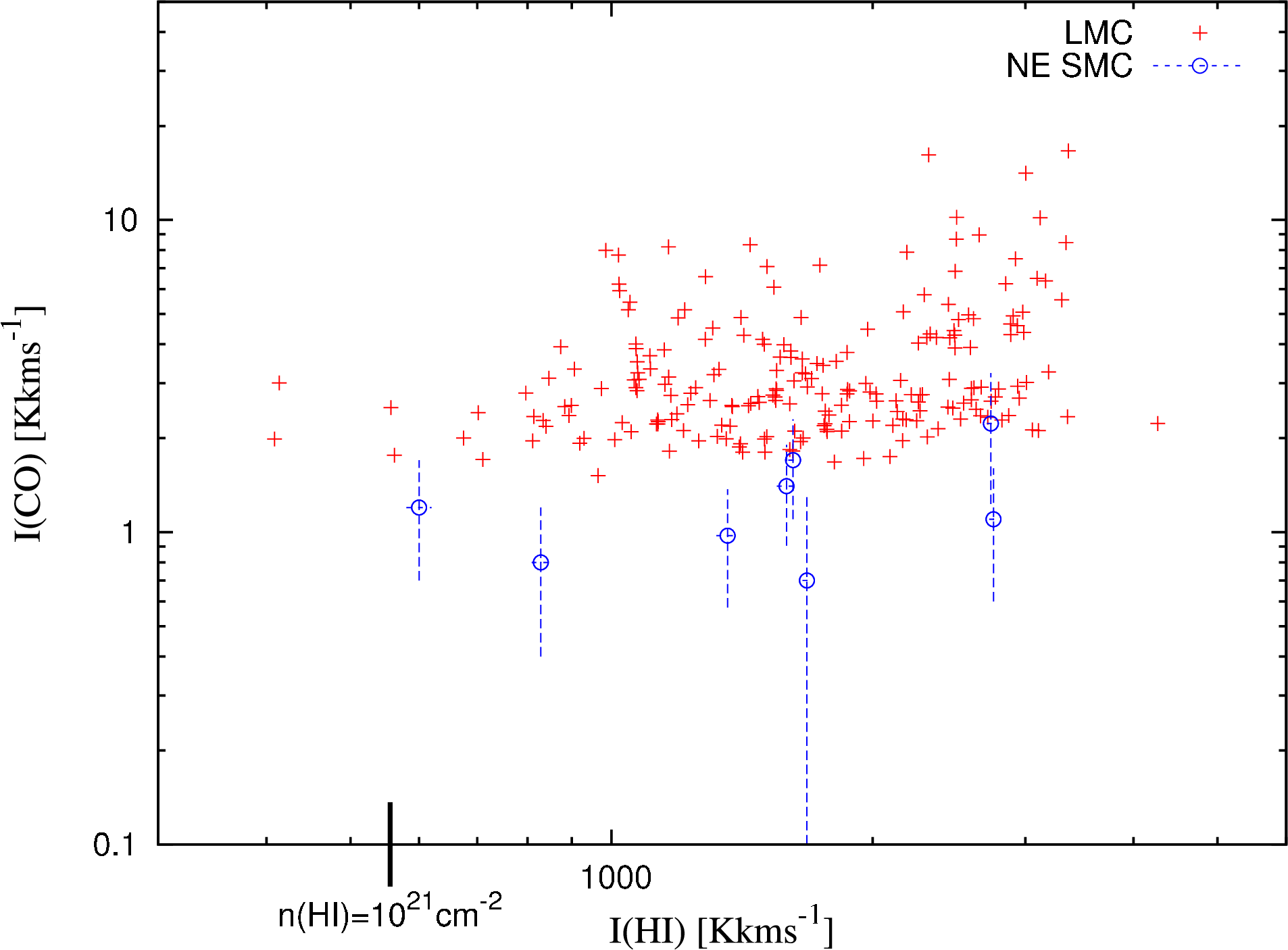}}}
\caption{\emph{Plus signs}: I$_{CO}$/I$_{HI}$ of Molecular clouds in the LMC \cite{wong}.\emph{Circles}: I$_{CO}$/I$_{HI}$ of Molecular clouds in the SMC. (This publication). CO clouds in the SMC are typically weaker than those in the LMC, and so the SMC data points plot  below those of the LMC. Note that the SMC Mopra data are smoothed to the 98'' resolution of the \hi\ dataset. An \hi\ column density of $n\approx$10$^{21}$  cm$^{-2}$ is also indicated with a vertical line, to indicate the $n$\hi\ star-forming threshold observed in the LMC by \cite{luks}\label{fig:cohicomp}
}
\end{figure}

We found in all cases that fitting two Gaussian components provided a significant improvement in the summed reduced $\chi^2$ (\srcs) parameter (by a factor of at least two, and a mean improvement by a factor of 4, calculated where T$_{\hi}>$3$\sigma_{\hi}$), but that fitting three components provided no further significant improvement. We have therefore fit two components to the \hi\ profile and constrained one of them according to the measured CO centroid velocity, as listed in Table~\ref{tab:1}. In general, constraining one of the two \hi\ components by the CO velocity centroids did not result in an large degradation of \srcs compared to a completely unconstrained two-component fit: in the unconstrained case, the \srcs\ was always of the order unity, while constraining the fit resulted in an average increase of the \srcs\ by 20\%, although an increase by 40\% was found for the most extreme cases. The largest \srcs\ value was returned for fitting to cloud NE-3c (\srcs=5.1) and the smallest \srcs\ value was returned from fitting to cloud NE-4a (\srcs=1.7). Note that in the case of NE-4, which contains two distinct \hi\ profiles, fits were made only to the component within the velocity range of interest. For example, while fitting to the high velocity component for NE-4c$_{high}$, the low-velocity \hi\ component was ignored. Constraining one of the fitted HI velocity centroids by the CO component usually resulted in a significant (up to 100\%) change in the integrated \hi\ from a completely unconstrained fit. Clouds NE-1a and NE-1c were the only examples where constraining the fit did not significantly change the measured integrated \hi. An example of this process applied  to cloud NE-3c is shown in Figure~\ref{fig:examplefit}, the integrated intensity of the \hi\ component associated with each CO velocity centroid is listed in Table~\ref{table:hicotable}, and plotted against CO integrated brightness in Figure~\ref{fig:cohicomp}. Figure~\ref{fig:cohicomp} also shows data from the study of the LMC by \cite{wong}. 

We do not have a means to presume velocity components to disentangle the complex \hi\ profiles at positions where CO is not detected. As such, we are unable to make a robust statistical examination of the null case; i.e the values of Integrated \hi\ where CO is not detected. Figure~\ref{fig:cohicomp} therefore presents only data-points where both CO and HI are detected, and we find that the ranges of these two qualities are consistent with those in the LMC. As we have observed some of the brightest CO regions in the northern SMC, this suggests that the conclusions made by \cite{wong} regarding the distribution of CO and \hi\ in the LMC apply also to the SMC: that CO in the northern SMC requires high \hi\ column densities as a minimum condition for formation. Note that the plot shows that the integrated intensities of the CO clouds in the SMC are generally smaller (by a few factors) than those of the LMC, which is an already-anticipated result \citep[e.g.][]{israel1993}. 
  
\begin{deluxetable}{|c|l|l|}

\tabletypesize{\scriptsize}
\tablecaption{I(CO) and I(HI) of northern CO cloud population \label{table:hicotable}}
\tablehead{
Cloud		& I$_{HI}$			&I$_{CO}$\\
		&[Kkms$^{-1}$]$\times$10$^3$	&[Kkms$^{-1}$]}
\startdata
NE-1a		& 1.62 $\pm$0.02		&1.7 $\pm$0.6\\
NE-1c		& 0.830 $\pm$0.02		&0.8 $\pm$0.4\\
NE-3a		& 1.68 $\pm$0.02		&0.7 $\pm$0.6\\
NE-3c		& 0.600 $\pm$0.02		&1.2 $\pm$0.5\\
NE-3f		& 2.76 $\pm$0.03		&1.1 $\pm$0.5\\
NE-4a		& 1.59 $\pm$0.04		&1.4 $\pm$0.5\\
NE-4c$_{hi}$	& 1.36 $\pm$0.04		&0.9 $\pm$0.4\\
NE-4c$_{low}$	& 2.74 $\pm$0.03		&2.2 $\pm$1
\enddata
\end{deluxetable}

\subsection{Dust, CO and \hi}
Figures~\ref{fig:ne23} B,D and~\ref{fig:ne4}C show the relative distribution of the detected CO against that of emission at \mum\ \citep{spitzer} for NE-1, NE-3 and NE-4 respectively. We defer a more comprehensive analysis of the relationship of the small-scale CO to the \mum\ emission to a later paper, however we generally find  from these new high-resolution data that the CO clouds in the more quiescent northern SMC are compact and are correlated only with the very brightest peaks of \mum\ emission; i.e. bright CO occurs only with bright \mum\ and \hi\ (e.g. the western CO peak in NE-3), and bright \mum\ occurs without any detectable CO (e.g. the eastern \mum\ peak in NE-3). 

Variations in the Dust-to-Gas ratio, temperature variations of the dusty component, or photo-disassociation processes may account for inconsistencies in the correlations of the \mum\ brightness and the CO integrated intensity but the point remains that the correlation of these emission datasets towards the northern SMC is not robust, emphasizing both the need for higher-resolution \hi\ observations, and also that estimates of the $X$ factor may be highly variable throughout the SMC \citep[see also][]{leroy07}.

\subsection{Estimates of $X_{CO}$}\label{sec:xco} 
Our high-resolution observations also allow us to make a new estimate of the SMC $X_{CO}$ factor using virial masses estimated directly from the data. Using the same approach as \cite{pineda}; $X\sim M_{vir}/L_{CO}$ (and assuming a ratio the total gas mass to H$_2$ gas mass of 1.36), we find a range of $X$ factor values of 5.4 $\times$10$^{20}$ - 4.3 $\times$10$^{21}$ cm$^{-2}$[K\kms]$^{-1}$, with a mean value of 1.3$^{+1.6}_{-0.8}\times$10$^{21}$ cm$^{-2}$[K\kms]$^{-1}$ for the northern SMC cloud population. The errors of these measurements are high due to the poor signal-to-noise ratio and matching of the beam and cloud radii (as discussed in Section~\ref{sec:3}). Note that this value is approximately a factor of four higher than that found for the LMC \citep{pineda} (after adjusting for their slightly different definition of M$_{vir}$).  Given that the ensemble virial masses (i.e. the velocity dispersions and radii) of the northern SMC population is approximately represented by the smallest members of the LMC cloud population \citep{pineda}, this suggests that the SMC harbors a significantly under-luminous CO-cloud population. 

The $X_{SMC}$ factor was also estimated by \cite{rubio1991} from low-resolution observations of the same northern population. Later $X_{SMC}$ factor estimates were made from higher-resolution observations of the south-west population in SKP-III. These two studies used different methods to estimate the $X_{SMC}$ factor: 
\noindent1. The earlier paper assumes that the size-line width relationship of CO clouds in the Galaxy and SMC is consistent, and they estimated that SMC CO clouds were under-luminous compared to the Galaxy by a factor of 20. The $X_{SMC}$ was therefore estimated to be $\sim$6$\times$10$^{21}$ cm$^{-2}$[K\kms]$^{-1}$; approximately 20 times higher than $X_{Gal}$. (using the values for $X_{Gal}$ at the time of their writing).
\noindent2. The later work by the SEST key projects using $\sim$45" resolution observations of clouds in the southwest of the SMC made estimates of the $X_{SMC}$ for each cloud by assuming a specific fractional mass of the molecular hydrogen. The resulting  $X_{SMC}$ values range between 2.5$\times$10$^{20}$ cm$^{-2}$[K\kms]$^{-1}$ and 2.6$\times$10$^{22}$ cm$^{-2}$[K\kms]$^{-1}$, with stronger clustering at approximately 10$^{21}$ cm$^{-2}$[K\kms]$^{-1}$. No errors were estimated for these values.

%We note that our measurements of the viral mass and luminosity of northern SMC clouds show no strong systematic deviation from the empirical relationship derived by \cite{rubio1993b} from the southwest population, (shown in Figure~\ref{fig:plots}, \emph{Left}). We can suggest then, that the $X_{SMC}$ factor derived by mm-observations is largely invariant between the northern and south-western regions.

The issue of the relative distributions of CO and H$_2$ was discussed in some detail by \cite{leroy07}, who used the 2'.6 FWHM NANTEN data to show that the CO was more compact than the H$_2$ (determined from 160\mum) by a factor of approximately 1.3. They also find that the  $X_{SMC}$ factor varies throughout the SMC by a factor of $\sim$2, however, using a factor of 1.3 as a parameter to adjust for the relative sizes of the H$_2$ and CO emission regions, they estimate a mean $X_{SMC}$ factor of 6$\times$10$^{21}$cm$^{-2}$[K\kms]$^{-1}$. We find from our higher-resolution data that the average radius of the clouds in the SMC is approximately 12 pc, which would result in a correction factor of approximately 3.3 and would adjust their FIR-based estimate of the $X_{SMC}$ to 2.4$\times$10$^{21}$cm$^{-2}$[K\kms]$^{-1}$. This value is consistent with CO-derived virial masses derived earlier. \cite{leroy07} indicate that the difference between the mm-derived virial masses and the $\mu$m-derived masses, which would indicate instability and would otherwise lead to cloud fragmentation, may be balanced by an ambient pervasive magnetic field. The small correction found from more exact cloud radii measurements presented here reduces the necessary strength of any such field \citep[see also][]{bot}. 

\section{summary}
We have produced the highest-resolution observations of the molecular cloud population in the northern part of the SMC, to a sensitivity of 210 mK per 0.9 \kms\ channel and resolving each cloud into 3-7 smaller clouds. We find that this northern population conforms to empirical luminosity-virial mass relationships found for the southern SMC population, yet they are generally much less luminous for their radius. Extrapolating the empirical size-linewidth relationship derived for the southern clouds implies that the northern population has minimum radii of a ~4 pc. A comparison with wider-field CO data suggests that two of the three cloud complexes do not exist in a large-scale and weak CO envelope, and that high-resolution \hi\ data are imperative for more clearly understanding the distribution of molecular material within the SMC. We find that, as for the LMC, CO in the SMC occurs only with high (i.e. $>$10$^{21}$ cm$^{-1}$) \hi\ column density, which suggests that as for the LMC, such column densities are required for its formation. We use these molecular data to estimate X$_{SMC}$ factors to be approximately  1.3$\times$10$^{21}$, and our high-resolution data allow us to apply a more appropriate correction factor to previous estimates of the X$_{SMC}$ from  $\mu$m+\hi\ data, and arrive at roughly consistent values.
%We also report that CO is detected at two closely-spaced lines of sight over two distinct velocity ranges, providing evidence for the action of some energetic process in the past evolution of the SMC which has resulted in the acceleration of the ISM at two separate velocity components.

\acknowledgments{}
All astronomical images used in this publication were made with the use of the {\sc karma} package \citep{karma}.  JLP was supported by an appointment to the NASA Postdoctoral Program at the Jet Propulsion Laboratory, California Institute of Technology, administered by Oak Ridge Associated Universities through a contract with NASA.
\bibliography{mybib}
\end{document}